\documentclass[twocolumn,prb,showpacs,citeautoscript,floatfix]{revtex4}
\usepackage{amsmath}
\usepackage{amssymb}
\usepackage{amsfonts}
\usepackage{graphicx}
\usepackage{dcolumn}
\usepackage{bm}
\usepackage{color}
\usepackage{chngcntr}
\usepackage{epstopdf}

\def\XXint#1#2#3{{\setbox0=\hbox{$#1{#2#3}{\int}$}
 \vcenter{\hbox{$#2#3$}}\kern-.5\wd0}}

\font\TenCM=cmr10
\font\SevenCM=cmr7
\font\FiveCM=cmr5
\newfam\CMfam

\textfont\CMfam=\TenCM
\scriptfont\CMfam=\SevenCM
\scriptscriptfont\CMfam=\FiveCM

\begin{document}

\title{Two-species Bose-Einstein condensate in an optical lattice:\\
analytical approximate formul\ae.}
\author{R.~Cipolatti}
\affiliation{Instituto de Matem\'atica, Universidade Federal do Rio de Janeiro, C.P.\
68530, Rio de Janeiro, RJ, Brazil}
\author{L.~Villegas-Lelovsky}
\affiliation{Instituto de F\'{\i}sica, Universidade de Bras{\'\i}lia,70.910-900, Bras{\'\i}lia, DF, Brazil}
\author{M.C.~Chung}
\affiliation{Department of Physics, National Chung Hsing University, Taichung, 40227, Taiwan}
\author{C.~Trallero-Giner}
\affiliation{Facultad de F\'{\i}sica, Universidad de La Habana, Vedado 10400, La Habana,
Cuba}
\date{\today }

\begin{abstract}
Employing a general variational method and perturbation theory, we derived
explicit solutions for the description of one-dimensional two species
Bose-Einstein condensates confined by a harmonic trap potential in an
optical lattice. We consider the system of two coupled Gross-Pitaevkii
equations (GPE) and derive explicit expressions for the chemical potentials
and wavefunctions in terms of the atom-atom interaction parameters and laser
intensity. We have compared our results with the numerical solutions of the
GPE and performed a quantitative analysis for the both considered methods.
We underline the importance of the obtained explicit solutions to
characterize the density profile or degree of miscibility of the two
components.
\end{abstract}

\maketitle

\section{Introduction}
\noindent
Multiple Bose-Einstein condensates (BEC) of different atomic species have
been realized in the last years. Mixture of alkali atoms of $^{87}$Rb in two
different hyperfine internal spin states~\cite{1}, atoms $^{23}$Na with a
superposition of spinor condensates~\cite{2}, combination of $^{41}$K - $%
^{87}$Rb,~\cite{3} $^{87}$Rb-$^{85}$Rb,~\cite{4} $^{87}Rb-^{133}$Cs,~\cite{5}
and gases of rare atomic species $^{168}$Yb-$^{174}$Yb,~\cite{6} have been
employed to produce two species BEC. These quantum degenerate mixtures allow
to study several intriguing phenomena as the dynamics of the superfluid
system,~\cite{2,3} the production of heteronuclear polar molecules,~\cite{7}
the miscibility or immiscibility of the two quantum fluids,~\cite{8} among
other effects. Also, two-species BEC loaded in a optical lattice have been
explored.~\cite{9,10} A similar system but of Fermi-Bose quantum gas mixture
in a 3-dimensional optical lattice was implemented to study the interfering
paths of the bosonic wave function scattered by the presence of fermionic
atoms.~\cite{11} These results have led to an intense theoretical and
mathematical studies on the properties of the two-coupled Gross-Pitaevkii
equations.

{The basis of this research lies on the knowledge of} the dependence of the chemical potentials as %
{functions} of the interparticle interactions and the spatial density probability.~\cite{13}

{A fascinating experimental realization to study the one dimensional (1D) transport
properties of ultracold fermionic and bosonic atoms in a periodic potential
have been reported in Ref.~\onlinecite{13}.}

From the theoretical point of view there are several studies for the
description of two species Bose condensates. Typically, numerical approaches
or Thomas-Fermi approximation are employed to calculate the chemical
potential and the ground state wave functions.~\cite{12} In Ref.~%
\onlinecite{14} it is analyzed the mixture of 1D two interacting condensates
modeled by the Bose-Hurbbard Hamiltonian and by using the quantum Monte
Carlo numerical simulations. Theoretical analysis of the 1D two component
BEC problem becomes an important reservoir to mimic different physical
effects of the Condensed Matter Physics (see for example Refs.~%
\onlinecite{15,16,17}), including the magnetic properties of the bosonic
mixtures with tunable interspecies interactions.~\cite{18} Also, as it will
be shown below, we can take advantage of analytical results for the study of
quantum effects and predictions for cold atoms researches.

Assuming a \textquotedblleft cigar-shaped\textquotedblright\ type for the
Bose-Einstein condensates\cite{Carretero,Trallero0} of a gas composed by two
kind of bosons loaded in an optical lattice, we can consider the following
system of 1D GP equations:

\begin{equation}
\mathbf{L}_{0}\mathbf{\Phi }+\left[ \mathbf{L}_{I}-\mathbf{\nu }\right]
\mathbf{\Phi }=\mathbf{0}~,  \label{Coup}
\end{equation}%
where $\mathbf{L}_{0}$ and $\mathbf{L}_{I}$ are, respectively the operators
\begin{equation}
\left[
\begin{array}{cc}
\displaystyle-\frac{\hbar ^{2}}{2m_{1}}\frac{d^{2}}{dx^{2}}+\frac{1}{2}%
m_{1}\omega_1^{2}x^{2} & 0 \\
0 & \displaystyle-\frac{\hbar ^{2}}{2m_{2}}\frac{d^{2}}{dx^{2}}+\frac{1}{2}%
m_{2}\omega_2^{2}x^{2}%
\end{array}%
\right] ~,  \label{B1}
\end{equation}%
\begin{equation}
\left[
\begin{array}{cc}
\overline{\lambda _{1}}\left\vert \Phi _{1}\right\vert ^{2}-V_{L}\cos
^{2}\left( \frac{2\pi x}{d}\right) & \overline{\lambda _{3}}\Phi _{1}\Phi
_{2} \\
\overline{\lambda _{3}}\Phi _{1}\Phi _{2} & \overline{\lambda _{2}}
\left\vert \Phi _{2}\right\vert ^{2}-V_{L}\cos ^{2}\left( \frac{2\pi x}{d}
\right)%
\end{array}
\right] ~,  \label{B2}
\end{equation}
and

\begin{equation}
\mathbf{\Phi }=\left[
\begin{array}{c}
\Phi _{1} \\
\Phi _{2}%
\end{array}%
\right] ~,\qquad \mathbf{\nu }=\left[
\begin{array}{c}
\nu _{1} \\
\nu _{2}%
\end{array}%
\right] ~.  \label{B3}
\end{equation}%
Here, $\omega_i >0$ denotes the harmonic trap frequencies where
for simplicity we consider the same for both condensates, i.e.,
$\omega_1=\omega_2=\omega$, $m_{i}>0$, and $\nu _{i}$ are, respectively, the mass and chemical
potential for the specie $i$ ($i=1$ and $2),$ $V_{L}>0$ and $d>0$ the
intensity and laser wavelength, $\overline{\lambda _{i}}$ takes into account
the self-interaction term for the $i$th specie, and $\overline{\lambda _{3}}$%
, the interaction between unlike particles of the species $1$ and 2. In this
system, the complex function $\Phi _{i}(x)$ is known~\cite{Carretero} as the
\textsl{macroscopic wavefunction or order parameter\/} of the $i$th
component and is defined as the expectation value of the corresponding field
operator, namely $\Phi _{i}(x)=\langle \widehat{\Phi }_{i}(x)\rangle $. The
functions $\Phi _{i}$ satisfy the normalization conditions
\begin{equation}
\int_{\mathbb{R}}|\Phi _{i}(x)|^{2}dx=N_{i}~,\,\,\,i=1,2~,
\label{RestNumElem}
\end{equation}%
where $N_{i}$ denotes the number of atoms of the $i$th specie.

It is worth to notice that in some situations, as in the case of \textsl{\
spinor condensates}, where one produces confinement of an atomic cloud of an
element in different spin states,~\cite{Carretero,saito} the condition (\ref%
{RestNumElem}) must be substituted by
\begin{equation*}
\int_{\mathbb{R}}|\Phi _{1}(x)|^{2}dx+\int_{\mathbb{R}}|\Phi
_{2}(x)|^{2}dx=N~,\quad N=N_{1}+N_{2}~.
\end{equation*}

We can rewrite the system (\ref{Coup}) in its dimensionless form, by
considering, for instance, $l=\sqrt{\hbar /(m_{1}\omega )}$, $x=l\xi $, and $%
\Phi _{i}(x)=\psi _{i}(\xi )/\sqrt{l}$, $i=1,2$, in which case we have
\begin{equation}
\mathcal{L}_{0}\mathbf{\Psi }+\bigl[\mathcal{L}_{I}-\mathbf{\mu }\bigr]
\mathbf{\Psi }={0}~,  \label{AdimCoup}
\end{equation}%
where $\mathcal{L}_{0}$ and $\mathcal{L}_{I}$ are respectively the operators
\begin{equation}
\left[
\begin{array}{cc}
\displaystyle-\frac{1}{2}\frac{d^{2}}{d\xi ^{2}}+\frac{1}{2}\xi ^{2} & 0 \\
0 & \displaystyle-\frac{a_{2}}{2}\frac{d^{2}}{d\xi ^{2}}+\frac{1}{2a_{2}}\xi
^{2}%
\end{array}
\right] ~,  \label{AdimB1}
\end{equation}%
\begin{equation}
\left[
\begin{array}{cc}
\displaystyle\lambda _{1}\left\vert \psi _{1}\right\vert
^{2}-V_{0}\cos^{2}(\alpha \xi ) & \lambda _{3}{\psi _{1}\psi _{2}} \\
\lambda _{3}{\psi _{1}\psi _{2}} & \displaystyle\lambda _{2}\left\vert \psi
_{2}\right\vert ^{2}-V_{0}\cos ^{2}(\alpha \xi )%
\end{array}
\right] ~,  \label{AdimB2}
\end{equation}
\begin{equation}
\mathbf{\Psi }=\left[
\begin{array}{c}
\psi _{1} \\
\psi _{2}%
\end{array}
\right] ~,\qquad \mathbf{\mu }=\left[
\begin{array}{c}
\mu _{1} \\
\mu _{2}%
\end{array}
\right] ~.  \label{AdimB3}
\end{equation}
Here, $a_{2}=m_{1}/m_{2}$, $\lambda _{i}=\overline{\lambda _{i}}/l\hbar
\omega $, ($i=1,2,3$), $V_{0}=V_{L}/\hbar \omega $, $\alpha =2l\pi /d$ and $%
\mu _{j}=\nu _{j}/\hbar \omega $ ($j=1,2$). For the system (\ref{AdimCoup}),
the energy functional can be cast as
\begin{eqnarray}
\mathbf{E}(\psi _{1},\psi _{2}) &=&E_{1}(\psi _{1})+E_{2}(\psi _{2})  \notag
\\
&&+\frac{\lambda _{3}}{2}\int_{R}|\psi _{1}(\xi )|^{2}|\psi _{2}(\xi
)|^{2}d\xi ~,  \label{DefEnergyE}
\end{eqnarray}%
with
\begin{eqnarray*}
E_{1}(\psi ) &=&\frac{1}{4}\int_{\mathbb{R}}\left\vert \psi ^{\prime }(\xi
)\right\vert ^{2}\,d\xi +\frac{1}{4}\int_{\mathbb{R}}\xi ^{2}|\psi (\xi
)|^{2}\,d\xi + \\
&&{}\frac{\lambda _{1}}{4}\int_{\mathbb{R}}|\psi (\xi )|^{4}\,d\xi -\frac{%
V_{0}}{2}\int_{\mathbb{R}}\cos ^{2}(\alpha \xi )|\psi (\xi )|^{2}\,d\xi ~, \\
E_{2}(\psi ) &=&\frac{a_{2}}{4}\int_{\mathbb{R}}\left\vert \psi ^{\prime
}(\xi )\right\vert ^{2}\,d\xi +\frac{1}{4a_{2}}\int_{\mathbb{R}}\xi
^{2}|\psi (\xi )|^{2}\,d\xi + \\
&&{}\frac{\lambda _{2}}{4}\int_{\mathbb{R}}|\psi (\xi )|^{4}\,d\xi -\frac{%
V_{0}}{2}\int_{\mathbb{R}}\cos ^{2}(\alpha \xi )|\psi (\xi )|^{2}\,d\xi ~.
\end{eqnarray*}%
Therefore, the partial Fr\'{e}chet derivatives of $\mathbf{E}$ are
\begin{eqnarray}
\partial _{1}\mathbf{E} &=&E_{1}^{\prime }(\psi _{1})+\lambda _{3}|\psi
_{2}(\xi )|^{2}\psi _{1}(\xi )~,  \label{Part_Diff} \\
\partial _{2}\mathbf{E} &=&E_{2}^{\prime }(\psi _{2})+\lambda _{3}|\psi
_{1}(\xi )|^{2}\psi _{2}(\xi )~.
\end{eqnarray}%
The minimum of the energy $\mathbf{E}(\psi _{1},\psi _{2})$ under the
restrictions $\int_{\mathbb{R}}|\psi _{i}(\xi )|^{2}\,d\xi =N_{i}$ satisfies
the Lagrange conditions for some constants $\mu _{i}/2$ $(i=1,2)$,
\begin{equation}
\partial _{1}\mathbf{E}=\mu _{1}\psi _{1}(\xi )~,\quad \partial _{2}\mathbf{E%
}=\mu _{2}\psi _{2}(\xi )~.  \label{sistema2bis}
\end{equation}%
Notice that (\ref{sistema2bis}) coincides with (\ref{AdimCoup}).

In previous works~\cite{PhysicaD, Trallero1, Trallero2, Trallero3}, we have
presented different methods to express the chemical potential $\mu $ and the
order parameter $\psi(\xi )$ as function of the interaction parameter $%
\lambda $ for the 1D Gross-Pitaevkii equation. In the present paper, we
adapt two of these methods (the generalized variational approach~\cite%
{PhysicaD} and perturbation theory) for the system (\ref{AdimCoup}), by
considering the vector chemical potential $\mathbf{\mu }$ as function of the
atom-atom interaction strength of each component $\lambda _{1}$, $\lambda
_{2}$ and the interaction between both species, $\lambda _{3}$.

The paper is organized as follows: in Section~II we present the mathematical
framework of the variational problem formulation, which characterizes the
condensate as ground state solution for the system (\ref{AdimCoup}), as well
as its equivalent integral representation. We also report an exact
representation of $\mu(\lambda_1,\lambda_2, \lambda_3)$ over which is based
our variational approach described in Section~III. In Section~IV we develop
the perturbation method valid for two coupled GP equations. Section~V is
devoted to present the results of these two approaches comparing with the
exact numerical solution of the system (\ref{AdimCoup}). Also, final
conclusions are delivered showing the range of validity of both considered
methods, with respect to parameter values employed for the description of
two-species Bose-Einstein condensate in an optical lattice.

\section{General mathematical framework}

In this section we establish the functional framework for the mathematical
analysis of existence, regularity and stability of ground state solutions
for the system (\ref{AdimCoup}). There is a great number of mathematical
work on these questions, some of them mentioned in the references below. The
eingenvalue problem (\ref{AdimCoup}) has an intrinsic mathematical interest,
but the ground state solutions (i.e., standing wave solutions of minimal
energy) play important role for condensates. By standing wave we mean
solution of the evolution equation
\begin{equation}
\,{\fam\CMfam\TenCM i\/}\frac{\partial \mathbf{\Psi }}{\partial t}=\bigl[%
\mathcal{L}_{0}+\mathcal{L}_{I}\bigr]\mathbf{\Psi }~,  \label{EvolAdimCoup}
\end{equation}%
of the form
\begin{equation*}
\mathbf{\Psi }(t,\xi )=\left[
\begin{array}{c}
\exp (-\,{\fam\CMfam\TenCM i\/}\mu _{1}t)\psi _{1}(\xi ) \\
\exp (-\,{\fam\CMfam\TenCM i\/}\mu _{2}t)\psi _{2}(\xi )%
\end{array}%
\right] ~.
\end{equation*}

\subsection{Existence of ground states and their stability}

We consider the following minimization problem
\begin{equation}
\mathbf{E}_{min}(\mathbf{\lambda })=\min \{\mathbf{E}(\mathbf{\Psi })\,;\,%
\mathbf{\Psi }\in \Sigma \} ,  \label{MinProb}
\end{equation}%
where $\lambda =(\lambda _{1},\lambda _{2},\lambda _{3})$, $\mathbf{\Psi }%
=(\psi_1 ,\psi_2 )$,
\begin{equation*}
\Sigma =\Bigl\{\mathbf{\Psi }\in\Xi\,;\,\int_{\mathbb{R}}|\psi_1
(\xi)|^{2}d\xi =N_{1},\,\,\int_{\mathbb{R}}|\psi_2(\xi)|^{2}d\xi=N_{2}\Bigr\}
\end{equation*}%
and $\Xi=\mathcal{V}\times \mathcal{V}$, where
\begin{equation*}
\mathcal{V}=\Bigl\{\psi\in H^1(\mathbb{R})\,;\, \int_{\mathbb{R}}\bigl[%
|\psi^{\prime}(\xi)|^2+\xi^2|\psi(\xi)|^2\bigr]\,d\xi<\infty\Bigr\}
\end{equation*}
and $H^1(\mathbb{R})$ is the standard Sobolev space.

Although the solutions of Eq.~(\ref{AdimCoup}) are in general complex valued
functions, we can restrict ourselves to just the real valued ones. This can
easily be justified because any solution of this system satisfies the
following inequality:~\cite{RolciOK} there exist $0<\delta \leq 1$ and $%
C(\delta )>0$ such that
\begin{equation}
|\mathbf{\Psi }(\xi )|^{2}+|\mathbf{\Psi }^{\prime }(\xi )|^{2}|\leq
C(\delta )\exp (-\delta \xi ^{2}),\quad \forall \xi \in \mathbb{R}.
\label{Decaimento}
\end{equation}

Indeed, assuming that
\begin{equation}  \label{ComplexSol}
\mathbf{\Psi}(\xi)=\left[
\begin{array}{c}
\psi_{1R}(\xi)+\,{\fam\CMfam\TenCM i\/}\psi _{1I}(\xi) \\
\psi _{2R}(\xi)+\,{\fam\CMfam\TenCM i\/}\psi _{2I})(\xi)%
\end{array}
\right]~,
\end{equation}
the exponential decay (\ref{Decaimento}) and a simple calculus gives
\begin{equation*}
\psi _{1R}^{\prime }\psi _{1I}-\psi _{1R}\psi _{1I}^{\prime } =\frac{d\hfil}{%
d\xi }\left( \frac{\psi _{1R}}{\psi _{1I}}\right)|\psi_{1I}|^2 =0~.
\end{equation*}
Therefore, $\psi_{1R}=\beta \psi _{1I}$ for some real constant $\beta \not=0$%
. The same holds for second component of $\mathbf{\Psi}$, which gives us $%
\psi _{2R}=\gamma \psi _{2I}$ for some constant $\gamma$. Hence, the
function
\begin{equation}
\mathbf{U}(\xi)=\left[
\begin{array}{c}
\sqrt{1+\beta^2}\psi_{1R}(\xi) \\
\sqrt{1+\gamma^2}\psi_{2R}(\xi)%
\end{array}%
\right]
\end{equation}
is a real solution of (\ref{AdimCoup}) and (\ref{ComplexSol}) is given by
\begin{equation}
\mathbf{\Psi}(\xi)=\left[
\begin{array}{c}
\displaystyle\left(\frac{1}{\sqrt{1+\beta^2}}+\frac{\,{\fam\CMfam\TenCM i\/}%
\beta}{\sqrt{1+\beta^2}}\right)\psi_{1R}(\xi) \\
\displaystyle\left(\frac{1}{\sqrt{1+\gamma^2}}+\frac{\,{\fam\CMfam\TenCM i\/}%
\gamma}{\sqrt{1+\gamma^2}}\right)\psi_{2R}(\xi)%
\end{array}%
\right]
\end{equation}

The existence of a minimal energy solution is a consequence of the
Gagliardo-Nirenberg inequality (see Theorem~1.3.7 in Ref.~%
\onlinecite{Thierry}), which in 1D allows us to show that the energy
functional $\mathbf{E}$ is bounded by bellow on the manifold $\Sigma $, for
all values of $\lambda _{i}\in \mathbb{R}$, $i=1,2,3$. With arguments of
convexity, we can show that the (real) solution of minimal energy is unique
provided that $\lambda_1$, $\lambda_2$ and $\lambda_3$ are positive.
Moreover, since the system (\ref{AdimCoup}) has the properties of
conservation of energy and mass (i.e., the number of particles), we can
prove the \textsl{orbital stability}~\cite{PhysicaD,Hajaiej} of ground
states.

On the other hand, the space $\mathcal{H}=L^2(\mathbb{R})\times L^2(\mathbb{R%
})$ is a Hilbert space if one considers the usual inner product
\begin{equation*}
(\mathbf{\Psi}|\mathbf{\Phi})_{\mathcal{H}} = \int_{\mathbb{R}%
}\psi_1(\xi)\phi_1(\xi)\,d\xi + \int_{\mathbb{R}}\psi_2(\xi)\phi_2(\xi)\,d\xi
\end{equation*}
and the differential operator
\begin{equation*}
\mathcal{L}_0:D(\mathcal{L}_0)\subset\mathcal{H}\rightarrow\mathcal{H}
\end{equation*}
is self-adjoint and maximal monotone\cite{Brezis}. So, it is invertible and
we can rewrite the equation (\ref{AdimCoup}) as
\begin{equation}  \label{AdimCoupIntegral}
\mathbf{\Psi} = \mathcal{L}_0^{-1}\bigl[\mu-\mathcal{L}_I\bigr]\mathbf{\Psi}.
\end{equation}
Since $D(\mathcal{L}_0) \subset\Xi$ and $\Xi$ is compactly embedded in $%
\mathcal{H}$\cite{PhysicaD}, $\mathcal{L}_0^{-1}$ is a compact integral
operator.

\subsection{Exact formul\ae }

We assume that, for each $\mathbf{\lambda }\in \mathbb{R}^{3}$, we can
choose $\mathbf{\Psi }_{\lambda }\in \Sigma $ such that the map $\mathbf{%
\lambda }\mapsto \mathbf{\Psi }_{\lambda }$ is a differentiable manifold in $%
\Xi$. Then, we have
\begin{eqnarray*}
{\frac{\partial \mathbf{E}_{min}}{\partial \lambda _{1}}}(\mathbf{\lambda })
&=&\Bigl\langle E_{1}^{\prime }(\psi _{1\mathbf{\lambda }}):{\frac{\partial
}{\partial \lambda _{1}}}\psi _{1\mathbf{\lambda }}\Bigr\rangle+{\frac{1}{4}}%
\Vert \psi _{1\mathbf{\lambda }}\Vert _{4}^{4} \\
&&{}+\Bigl\langle E_{2}^{\prime }(\psi _{2\mathbf{\lambda }}):{\frac{%
\partial }{\partial \lambda _{1}}}\psi _{2\mathbf{\lambda }}\Bigr\rangle \\
&&{}+\lambda _{3}\Bigl\langle|\psi _{2\mathbf{\lambda }}|^{2}\psi _{1\mathbf{%
\lambda }}:{\frac{\partial }{\partial \lambda _{1}}}\psi _{1\mathbf{\lambda }%
}\Bigr\rangle \\
&&{}+\lambda _{3}\Bigl\langle|\psi _{1\mathbf{\lambda }}\mathbf{|}^{2}\psi
_{2\mathbf{\lambda }}:{\frac{\partial }{\partial \lambda _{1}}}\psi _{2%
\mathbf{\lambda }}\Bigr\rangle \\
&=&\mu _{1}(\mathbf{\lambda )}\frac{\partial \hfil}{\partial \lambda _{1}}%
\Vert \psi _{1\mathbf{\lambda }}\Vert _{2}^{2}+\mu _{2}(\mathbf{\lambda }){%
\frac{\partial \hfil}{\partial \lambda _{1}}}\Vert \psi _{2\mathbf{\lambda }%
}\Vert _{2}^{2} \\
&&{}+{\frac{1}{4}}\Vert \psi _{1\mathbf{\lambda }}\Vert _{4}^{4}~.
\end{eqnarray*}%
Since $(\psi _{1\lambda },\psi _{2\lambda })\in \Sigma $ implies
\begin{equation*}
{\frac{\partial \hfil}{\partial \lambda _{1}}}\Vert \psi _{1\mathbf{\lambda }%
}\Vert _{2}^{2}={\frac{\partial \hfil}{\partial \lambda _{1}}}\Vert \psi _{2%
\mathbf{\lambda }}\Vert _{2}^{2}=0~,
\end{equation*}%
we get
\begin{equation*}
{\frac{\partial \mathbf{E}_{min}}{\partial \lambda _{1}}}(\mathbf{\lambda })=%
{\frac{1}{4}}\Vert \psi _{1\mathbf{\lambda }}\Vert _{4}^{4}~.
\end{equation*}%
\textsl{Mutatis-mutandis\/}, we have
\begin{equation*}
{\frac{\partial \mathbf{E}_{min}}{\partial \lambda _{2}}}(\mathbf{\lambda })=%
{\frac{1}{4}}\Vert \psi _{2\mathbf{\lambda }}\Vert _{4}^{4}
\end{equation*}%
and with the same arguments, we obtain
\begin{equation*}
{\frac{\partial \mathbf{E}_{min}}{\partial \lambda _{3}}}(\mathbf{\lambda })=%
{\frac{1}{2}}\Vert \psi _{1\mathbf{\lambda }}\psi _{2\mathbf{\lambda }}\Vert
_{2}^{2}~.
\end{equation*}%
Therefore,
\begin{equation*}
\nabla \mathbf{E}_{min}(\mathbf{\lambda })=\left( {\frac{1}{4}}\Vert \psi _{1%
\mathbf{\lambda }}\Vert _{4}^{4},{\frac{1}{4}}\Vert \psi _{2\mathbf{\lambda }%
}\Vert _{4}^{4},\frac{1}{2}\Vert \psi _{1\mathbf{\lambda }}\psi _{2\mathbf{%
\lambda }}\Vert _{2}^{2}\right)
\end{equation*}
and
\begin{equation*}
\mathbf{E}_{min}(\mathbf{\lambda })=\mathbf{E}_{min}(0)+\int_{0}^{1}\nabla
\mathbf{E}_{min}(\mathbf{\lambda }(s))\cdot {\frac{d}{ds}}\mathbf{\lambda }%
(s)\,ds~,
\end{equation*}
for any smooth path $\mathbf{\lambda }(s)$ in $\mathbb{R}^{3}$ joining the
points $(0,0,0)$ and $(\lambda _{1},\lambda _{2},\lambda _{3})$. In
particular, for the linear path $\mathbf{\lambda }(s)=s\mathbf{\lambda }%
=(s\lambda _{1},s\lambda _{2},s\lambda _{3})$, $0\leq s\leq 1$, for which we
have the following formula

\begin{equation}
\begin{array}{rcl}
\mathbf{E}_{min}(\mathbf{\lambda }) & = & \mathbf{E}_{min}(0,0,0)%
\displaystyle{}+\frac{1}{4}\int_{0}^{1} \Bigl(\Vert \psi _{s1\mathbf{\lambda
}}\Vert _{4}^{4}\lambda _{1} \\
&  & {}+\Vert \psi _{s2\mathbf{\lambda }}\Vert_{4}^{4}\lambda _{2} +2\Vert
\psi _{s1\mathbf{\lambda }}\psi _{s2\mathbf{\lambda }} \Vert _{2}^{2}\lambda
_{3}\Bigr)\,ds~.%
\end{array}
\label{EnerMin}
\end{equation}

The chemical potentials $\mu _{1}$ and $\mu _{2}$ as function of the
parameter $\mathbf{\lambda }$ can be easily calculated by multiplying the
first equation of (\ref{AdimCoup}) by $\psi_{1\mathbf{\lambda }}$, the
second by $\psi_{2\mathbf{\lambda }}$ and taking the integral over $\mathbb{%
R }$. By this calculation, we get

\begin{eqnarray}
\mu _{1}(\lambda ) &=&\frac{2}{N_{1}}\left( E_{1}(\psi _{1\lambda })+\frac{
\lambda _{1}}{4}\Vert \psi _{1\lambda }\Vert _{4}^{4}+\frac{\lambda _{3}}{2}
\Vert \psi _{1\lambda }\psi _{2\lambda }\Vert _{2}^{2}\right) ~,  \notag \\
&&  \label{mu1} \\
\mu _{2}(\lambda ) &=&\frac{2}{N_{2}}\left( E_{2}(\psi _{2\lambda })+\frac{
\lambda _{2}}{4}\Vert \psi _{2\lambda }\Vert _{4}^{4}+\frac{\lambda _{3}}{2}
\Vert \psi _{1\lambda }\psi _{2\lambda }\Vert _{2}^{2}\right) ~.  \notag \\
&&  \label{mu2}
\end{eqnarray}

\section{Variational approach}

We consider the following trial functions:
\begin{equation}
\psi _{k}(\xi )=\sqrt{N_{k}}\left( \frac{2\tau _{k}}{\pi }\right) ^{1/4}\exp
(-\tau _{k}\xi ^{2})~,\,\,k=1,2~.  \label{TrialFunc}
\end{equation}%
By calculating the energy $\mathbf{E}$ with these functions, we get:
\begin{eqnarray*}
\mathbf{E}(\psi _{1},\psi _{2}) &=&\displaystyle\sum_{k=1}^{2}N_{k}\left(
\frac{a_{k}}{4}\tau _{k}+\frac{1}{16 a_{k}\tau _{k}}+\frac{\lambda
_{k}N_{k}\tau _{k}^{1/2}}{4\sqrt{\pi }}\right. \\
&&\displaystyle\left. {}-\frac{V_{0}}{4}\left( 1+e^{-\alpha ^{2}/2\tau
_{k}}\right) \right) \\
&&\displaystyle\qquad {}+\frac{\lambda _{3}N_{1}N_{2}}{\sqrt{2\pi }}\left( {%
\frac{\tau _{1}\tau _{2}}{\tau _{1}+\tau _{2}}}\right) ^{1/2}~.
\end{eqnarray*}%
where, to simplify the notation, we introduced $a_{1}=1$. So, by denoting $%
f(\tau _{1},\tau _{2})=\mathbf{E}(\psi _{1},\psi _{2})$, it is easy to see
that $f(\tau _{1},\tau _{2})$ is bounded by bellow. Indeed, if $\lambda
_{3}\geq 0$, we have
\begin{eqnarray*}
f(\tau _{1},\tau _{2}) &\geq &\sum_{k=1}^{2}N_{k}\left( \frac{a_{k}}{4}\tau
_{k}+\frac{1}{16 a_{k}\tau _{k}}+\frac{\lambda _{k}N_{k}}{4\sqrt{\pi }}\tau
_{k}^{1/2}\right. \\
&&\left. \quad {}-\frac{V_{0}}{4}\left( 1+e^{-\alpha ^{2}/2\tau _{k}}\right)
\right)
\end{eqnarray*}
and the conclusion is evident. Otherwise, notice that
\begin{eqnarray*}
\left( {\frac{\tau _{1}\tau _{2}}{\tau _{1}+\tau _{2}}}\right) ^{1/2} &\leq &%
{\frac{1}{2}}\sqrt{\tau _{1}+\tau _{2}}\leq {\frac{\sqrt{2}}{2}}\max \bigl\{%
\sqrt{\tau _{1}},\sqrt{\tau _{2}}\bigr\} \\
&\leq &{\frac{\sqrt{2}}{2}}(\sqrt{\tau _{1}}+\sqrt{\tau _{2}})~,\quad
\forall \,\tau _{1},\,\tau _{2}>0
\end{eqnarray*}%
from which we get
\begin{eqnarray*}
f(\tau _{1},\tau _{2}) &\geq &\sum_{k=1}^{2}N_{k}\left[ \frac{a_{k}}{4}\tau
_{k}+\frac{1}{16 a_{k}\tau _{k}}\right. \\
&&\quad {}+{\frac{1}{2\sqrt{\pi }}}\left( \frac{\lambda _{k}N_{k}}{2}+\frac{%
\lambda _{3}N_{1}N_{2}}{N_{k}}\right) \sqrt{\tau _{k}} \\
&&\qquad \left. {}-\frac{V_{0}}{4}\left( 1+e^{-\alpha ^{2}/2\tau
_{k}}\right) \right] ~.
\end{eqnarray*}
Hence, $f(\tau _{1},\tau _{2})$ riches its minimum at some $\tau
_{k}(\lambda _{1},\lambda _{2},\lambda _{3})$, ($k=1,2$) which are
necessarily solutions of the algebraic system ($i\not=j$):

\begin{multline}
a_{i}^{2}\tau _{i}^{2}+\frac{\lambda _{i}N_{i}a_{i}} {2\sqrt{ \pi }}\tau
_{i}^{3/2}+\frac{2\lambda _{3}N_{1}N_{2}a_{i}}{\sqrt{2\pi }N_{i}}\left( {%
\frac{\tau _{i}\tau _{j}}{\tau _{i}+\tau _{j}}}\right) ^{3/2}
\label{Sist_Algeb1} \\
-\frac{V_{0}\alpha ^{2}a_{i}}{2}e^{-\alpha ^{2}/2\tau _{i}}=\frac{1}{4}\text{
},\quad i\not=j~.
\end{multline}%
These are the equations to be solved in order to obtain $\tau _{1}(\mathbf{%
\lambda })$ and $\tau _{2}(\mathbf{\lambda })$ which will be used in the
formulas of $\mu _{app,1}(\mathbf{\lambda })$ and $\mu _{app,2}(\mathbf{%
\lambda })$ (see below). Notice that if $\lambda _{1}\not=\lambda _{2}$, the
respective roots are different even in the case $V_{0}=0,$ $N_{1}=N_{2}=N$
and $m_{1}=m_{2}$. Indeed, by subtracting the first equation from the second
one in (\ref{Sist_Algeb1}), we obtain:
\begin{equation*}
\tau _{2}^{2}-\tau _{1}^{2}={\frac{N}{2\sqrt{\pi }}}\bigl[\lambda _{1}\tau
_{1}^{3/2}-\lambda _{2}\tau _{2}^{3/2}\bigr]~,
\end{equation*}%
and we see that, if $\tau _{1}=\tau _{2}$, then $\lambda _{1}\tau
_{1}^{3/2}-\lambda _{2}\tau _{2}^{3/2}=0$, which implies that $\lambda
_{1}=\lambda _{2}$.

By choosing, $\sigma _{i}=\sqrt{a_{i}\tau _{i}}$, $i=1,2$, the equations (%
\ref{Sist_Algeb1}) can be written as
\begin{multline}
\sigma _{i}^{4}+\frac{\Lambda _{i}}{2\sqrt{\pi }}\sigma _{i}^{3}+\frac{\sqrt{%
2}\Lambda _{ij}a_{i}\sqrt{{a_j}}}{\sqrt{\pi }} \left( {\frac{\sigma
_{i}^{2}\sigma _{j}^{2}}{a_{j}\sigma _{i}^{2}+a_{i}\sigma _{j}^{2}}}\right)
^{3/2}  \label{Sist_Algeb2} \\
-\frac{V_{0}\alpha ^{2}a_{i}}{2}e^{-\alpha ^{2}a_{i}/2\sigma _{i}^{2}}={%
\frac{1}{4}}~,\quad i\not=j~,
\end{multline}%
where, for $i\not=j$, $\Lambda _{i}=\lambda _{i}N_{i}/\sqrt{a_{i}}$ and $%
\Lambda _{ij}=\lambda _{3}N_{j}/\sqrt{a_{j}}$.

\subsection{Approximate formul\ae }

Let $\sigma _{1}(\Lambda )$ and $\sigma _{2}(\Lambda )$ with $\Lambda
=(\Lambda _{1},\Lambda _{2},\Lambda _{12},\Lambda _{21}),$ the solution of
the system (\ref{Sist_Algeb2}).

Using Eqs.~(\ref{mu1}) and (\ref{mu2}), a direct calculation gives:
\begin{eqnarray}
\mu _{app,1}(\Lambda ) &=&\frac{\sigma _{1}^{2}(\Lambda )}{2}+{\frac{1}{%
8\sigma _{1}^{2}(\Lambda )}}+{\frac{\Lambda _{1}}{\sqrt{\pi }}}\sigma
_{1}(\Lambda )  \notag \\
&&\quad {}-{\frac{V_{0}}{2}}\left( 1+e^{-\alpha ^{2}/2\sigma
_{1}^{2}(\Lambda )}\right)  \notag \\
&&\qquad {}+{\frac{\sqrt{2}\Lambda _{12} \sqrt{a_{2}}}{\sqrt{\pi }}}\left( {%
\frac{\sigma _{1}^{2}(\Lambda )\sigma _{2}^{2}(\Lambda )}{a_{2}\sigma
_{1}^{2}(\Lambda )+\sigma _{2}^{2}(\Lambda )}}\right) ^{1/2}~,  \notag \\
&&  \label{muapr1} \\
\mu _{app,2}(\Lambda ) &=&\frac{\sigma _{2}^{2}(\Lambda )}{2}+{\frac{1}{%
8\sigma _{2}^{2}(\Lambda )}}+{\frac{\Lambda _{2}}{\sqrt{\pi }}}\sigma
_{2}(\Lambda )  \notag \\
&&\quad {}-{\frac{V_{0}}{2}}\left( 1+e^{-\alpha ^{2}a_{2}/2\sigma
_{2}^{2}(\Lambda )}\right) .  \notag \\
&&\qquad {}+{\frac{\sqrt{2}\Lambda _{21}}{\sqrt{\pi }}}\left( {\frac{\sigma
_{1}^{2}(\Lambda )\sigma _{2}^{2}(\Lambda )}{a_{2}\sigma _{1}^{2}(\Lambda
)+\sigma _{2}^{2}(\Lambda )}}\right) ^{1/2}~.  \notag \\
&&  \label{muapr2}
\end{eqnarray}

\subsection{Properties of the wavefunction and the minimal energy}

\begin{figure}[tbp]
\includegraphics[width = 0.48\textwidth]{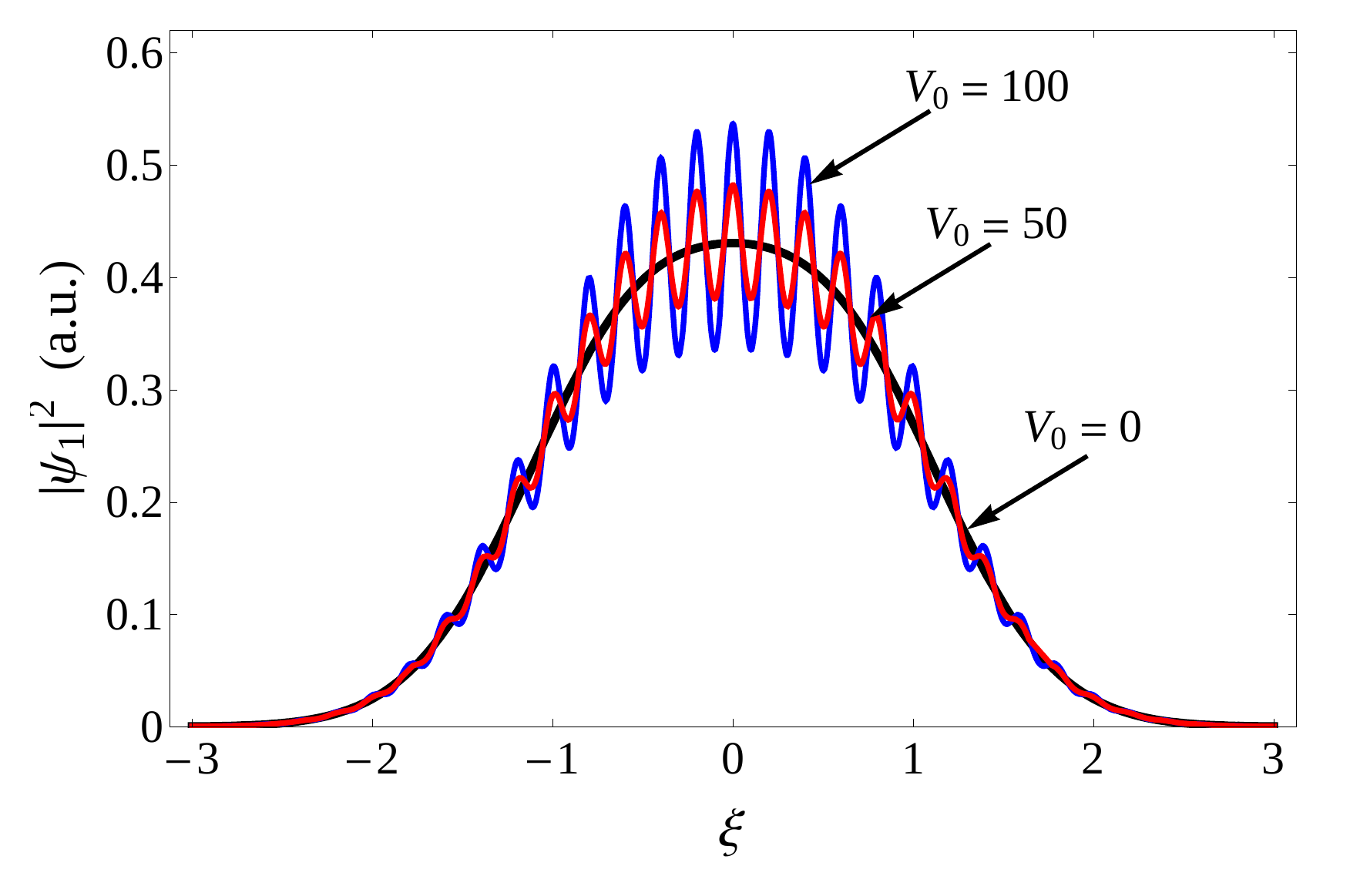}
\caption{(Color online) Normalized density probability for the order
parameter of one specie, $|\protect\psi _{1}(\protect\xi )|^{2}$ ($\Lambda
_{3}=0$), for $\Lambda _{1}=2$, $d/l=0.4$ and values of the laser intensity $%
V_{0}=0,50, $ and 100. Solution of $\protect\psi _{1}(\protect\xi )$ taken
from Ref.~ \onlinecite{Trallero3}.}
\label{fig:fig0}
\end{figure}

As it was pointed out in Eq.~(\ref{Decaimento}), each component of $\mathbf{%
\ \Psi }(\xi )$ in Eq.~(\ref{AdimB3}) behaves as a Gaussian as $\xi
\rightarrow \pm \infty $, for all values of $\Lambda $, $\mu $ and $V_{0}$.
In a general way, this behaviour justify the selection of the trial function
(\ref{TrialFunc}). Nevertheless, as it is achieved in Fig.~(\ref{fig:fig0}),
the variation of the wavefunction of one specie with respect to the optical
lattice intensity, $V_{0}=V_{L}/\hbar \omega $ and the reduced wavelength, $%
\alpha ^{-1}=d/(2l\pi )$ cannot be accounted by a Gaussian trial function (%
\ref{TrialFunc}). The strong variation of the optical lattice potential $%
U(\xi )=-V_{0}\cos ^{2}(\alpha \xi )$ with respect to $\alpha $ and $V_{0},$
keep off the contribution of the monotonic behavior of the harmonic
potential $\xi ^{2}$ to order parameter. Thus, the variational approach
presented here does not allow good results in the case $V_{0}\not=0$ is
large enough. Indeed, by an effective numerical solution of the 1D
Gross-Pitaevskii equation we obtain the order parameter $\psi (\xi )$ as
shown in Figure~\ref{fig:fig0}. On the other hand, if we consider the
equivalent formula of (\ref{Sist_Algeb1}) for the one component BEC, we
obtain\cite{PhysicaD}
\begin{equation}
\sigma ^{4}+\frac{\lambda }{4\sqrt{\pi }}\sigma ^{3}-\frac{V_{0}\alpha ^{2}}{%
4}e^{-\alpha ^{2}/2\sigma ^{2}}=\frac{1}{4}.  \label{DepV0}
\end{equation}%
For $\lambda \geq 0$ fixed, the function $\sigma (V_{0})$ implicitly defined
by Eq.~(\ref{DepV0}) satisfies the differential equation
\begin{equation}
\frac{d\sigma }{dV_{0}}=\frac{\alpha ^{2}e^{-\alpha ^{2}/2\sigma ^{2}}}{%
16\sigma ^{3}+\frac{3\lambda }{\sqrt{\pi }}\sigma ^{2}-\frac{V_{0}\alpha ^{4}%
}{\sigma ^{3}}e^{-\alpha ^{2}/2\sigma ^{2}}}  \label{EDO2V0}
\end{equation}%
which shows that it is increasing and blows up for a certain $V_{0}$ large
enough.

The choice of a test function that takes into account the variation shown in
the figure will be treated in a future publication.

Also, the presence of two-species introduces an effective interaction of the
unlike particle, which is described in our model by the coefficient $\Lambda
_{3}$. The effect of the $\Lambda _{3}\left\vert \psi _{j}\right\vert
^{2}\psi _{i}$ term on the condensates is to attract ($\lambda _{3}<0$) or
to repel ($\lambda _{3}>0$) the cloud probability densities $\left\vert \psi
_{i}(\xi )\right\vert ^{2}.$ Thus, in the case we are dealing with a strong
repulsive interaction, the maximum of the density probability lies at $\xi
\neq 0.$ Notice that the nature of our trial functions does not take into
account the present peculiarity of two-species BEC. In Sec.~V below we
present a brief discussion of this effect.

\section{Perturbation theory}

Following the result of Eq.~(\ref{AdimCoupIntegral}), we can write the
system of coupled integral equations
\begin{equation}
\mathbf{\Psi }(\xi )=\int_{-\infty }^{\infty }\mathbf{G}(\xi ,\xi ^{\prime })%
\left[ \mathbf{\mu -}\mathcal{L}_{I}\right] \mathbf{\Psi }(\xi ^{\prime
})d\xi ^{\prime }~,  \label{Inte2}
\end{equation}%
where the kernel
\begin{equation}
\mathbf{G}(\xi ,\xi ^{\prime })=\left(
\begin{tabular}{cc}
$G_{1}(\xi ,\xi ^{\prime })$ & 0 \\
0 & $G_{2}(\xi ,\xi ^{\prime })$%
\end{tabular}%
\right) ~,  \label{G}
\end{equation}%
is the solution of the differential equations $\mathcal{L}_{0}\mathbf{G}(\xi
,\xi ^{\prime })=I\delta (\xi -\xi ^{\prime })$ and $I$ the identity matrix.
In the spectral representation we have the Green function~\cite{morse}
\begin{equation}
G_{i}(\xi ,\xi ^{\prime })=\sum_{n=0}^{\infty }\frac{\varphi _{n}(\xi
/l_{i})\varphi _{n}(\xi ^{\prime }/l_{i})}{(n+1/2)}\text{ \ \ \ },\ i=1,2~.
\label{Gri}
\end{equation}%
with $l_{i}=\sqrt{a_{i}}$ and $\varphi _{n}(z)$ is the harmonic oscillator
wavefunction.~\cite{Landau}\ Thus, inserting $\mathbf{G}(\xi ,\xi ^{\prime
}) $ in (\ref{Inte2}) we get
\begin{equation}
\mathbf{\Psi }(\xi )=\left[
\begin{array}{c}
\psi _{1}(\xi ) \\
\psi _{2}(\xi )%
\end{array}%
\right] =\sum_{n=0}^{\infty }\left[
\begin{array}{c}
\sqrt{N_{1}}C_{n}\varphi _{n}(\xi /l_{1}) \\
\sqrt{N_{2}}D_{n}\varphi _{n}(\xi /l_{2})%
\end{array}%
\right] ~,  \label{fhino}
\end{equation}%
where the vectors $\mathbf{C=(}C_{1},C_{2},....\mathbf{)}$ and $\mathbf{D=(}%
D_{1},D_{2},....\mathbf{)}$ are given by
\begin{multline}
C_{n}[D_{n}]=\frac{1}{(n+\frac{1}{2})}\int\limits_{-\infty }^{\infty }\left[
\left( \mu _{1[2]}-\displaystyle\lambda _{1[2]}\left\vert \psi _{1[2]}(\xi
^{\prime })\right\vert ^{2}\right. \right.  \label{cn} \\
\left. +V_{0}\cos ^{2}(\alpha \xi ^{\prime })\right) + \\
\left. \lambda _{3}\left\vert \psi _{2[1]}(\xi ^{\prime })\right\vert ^{2}
\right] \varphi _{n}(\xi ^{\prime }/l_{1[2]})\psi _{1[2]}(\xi ^{\prime
})d\xi ^{\prime }~,
\end{multline}

To satisfy Eqs.~(\ref{fhino}) and (\ref{cn}), the vector coefficients $%
\mathbf{C}$ and $\mathbf{D}$ must fulfill the non-linear system of equations
\begin{eqnarray}
0 &=&\left[ \mathbf{\Delta }^{\mathbf{(1)}}(\mu _{1})+\Lambda _{1}\mathbf{%
C\cdot T}\cdot \mathbf{C}+{}\right.  \notag \\
&&\left. \Lambda _{12}\mathbf{D\cdot S(}l_{r}\mathbf{)}\cdot \mathbf{D}-V_{0}%
\mathbf{P}(\alpha \sqrt{a_{1}})\right] \mathbf{C~,}  \label{delta1}
\end{eqnarray}%
\begin{eqnarray}
0 &=&\left[ \mathbf{\Delta }^{\mathbf{(2)}}(\mu _{2})+\Lambda _{2}\mathbf{%
D\cdot T}\cdot \mathbf{D}+{}\right.  \notag \\
&&\left. \Lambda _{21}\mathbf{C\cdot S(}\frac{1}{l_{r}}\mathbf{)}\cdot
\mathbf{C}-V_{0}\mathbf{P}(\alpha \sqrt{a_{2}})\right] \mathbf{D}~,
\label{delta2}
\end{eqnarray}%
where $l_{r}=\sqrt{a_{1}/a_{2}},$ $\mathbf{\Delta }_{nm}^{\mathbf{(}i\mathbf{%
)}}=\left( n+1/2-\mu _{i}\right) \delta _{nm},$ $\mathbf{T}$ and $\mathbf{P}%
(\alpha )$ are matrices given elsewhere~\cite{Trallero3} and $\mathbf{S}(z)$
is defined in the Appendix A.

The above system is an infinite generalized eigenvalue problem for $\mu _{i}$
$(i=1,2)$\ and the vector coefficients $\mathbf{C}$ and $\mathbf{D}$. An
efficient algorithm for solving Eqs.~(\ref{delta1})-(\ref{delta2}) is
presented in Ref.~\onlinecite{Trallero2}. Nevertheless, it is very useful to carry
with explicit expressions for $\mu _{i}$ and $\psi _{i}$ in terms of the
leading parameters $\mathbf{\Lambda }$ and $V_{0}.$  Assuming that the
contribution of the non-linear terms and the optical potential appearing in
the system (\ref{delta1})-(\ref{delta2}) are small enough in comparison with
that of the harmonic potentials, allows that the vector solutions $\mathbf{%
\mu }$, $\mathbf{C}$ and $\mathbf{D}$ can be sought as Taylor polynomials of
the parameters $\mathbf{\Lambda }$ and $V_{0}$. Up to second order terms,
and solving simultaneously the system (\ref{delta1})-(\ref{delta2}), it is
possible to show that the chemical potentials is given by
\begin{multline}
\mu _{per,1}=\frac{1}{2}+\frac{\Lambda _{1}}{\sqrt{2\pi }}+\frac{\Lambda
_{12}}{\sqrt{\pi (1+l_{r}^{2})}}-\frac{V_{0}}{2}\left[ 1+\exp \left( -\alpha
^{2}a_{1}\right) \right]  \label{perturb} \\
-0.033106\Lambda _{1}^{2}.+\frac{\sqrt{2}\Lambda _{12}}{\pi \sqrt{1+l_{r}^{2}%
}}\left[ 2\Lambda _{1}f\left( l_{r}\right) +\Lambda _{2}f\left( \frac{1}{%
l_{r}}\right) \right] \\
+\frac{\Lambda _{12}}{\pi (1+l_{r}^{2})}\left[ \Lambda _{12}g\left(
l_{r}\right) +2l_{r}\Lambda _{21}g\left( \frac{1}{l_{r}}\right) \right] + \\
\frac{\exp \left( -\alpha ^{2}a_{1}\right) V_{0}}{\sqrt{2\pi }}\left[ \frac{%
\Lambda _{12}}{\sqrt{2(1+l_{r}^{2})}}h\left( \frac{l_{r}^{2}\alpha ^{2}a_{1}%
}{1+l_{r}^{2}}\right) \right. + \\
\left. \Lambda _{1}h\left( \frac{\alpha ^{2}a_{1}}{2}\right) \right] +\frac{%
\exp \left( -\alpha ^{2}a_{2}\right) V_{0}\Lambda _{12}}{\sqrt{2\pi
(1+l_{r}^{2})}}h\left( \frac{\alpha ^{2}a_{2}}{1+l_{r}^{2}}\right) \\
-\frac{\exp \left( -2\alpha ^{2}a_{1}\right) V_{0}^{2}}{4}ch(2\alpha
^{2}a_{1})~.
\end{multline}%
Functions $f(z),$ $g(z)$ and $h\left( z\right) $ and $ch(z)$ are defined in
Appendix B.

Finally, the dimensionless order parameter, $\psi _{1},$ considering
corrections up to the first order in $\Lambda _{1},$ $\Lambda _{12},$ and $%
V_{0}$, can be expressed as
\begin{eqnarray}
\psi _{per,1} &=&\varphi _{_{0}}(\xi )+\sum_{m=1}^{\infty }\left\{ \frac{%
(-1)^{m+1}\sqrt{(2m)!}}{\sqrt{\pi }2^{m}(m!)2m}\left[ \frac{\Lambda _{1}}{%
2^{m}\sqrt{2}}\right. \right.  \notag \\
&&\left. +\frac{\Lambda _{12}}{\sqrt{1+l_{r}^{2}}}\left( \frac{l_{r}^{2}}{%
1+l_{r}^{2}}\right) ^{m}\right] +V_{0}\frac{(-1)^{m}2^{m-1}}{\sqrt{(2m!)}}
\notag \\
&&\left. \times \left( \alpha ^{2}a_{1}\right) ^{m}\exp \left( -\alpha
^{2}a_{1}\right) \right\} \varphi _{2m}(\xi )~.  \label{fiper}
\end{eqnarray}%
The series, appearing in Eq.~(\ref{fiper}), can be summed obtaining the
compact solution
\begin{eqnarray}
\psi _{per,1} &=&\varphi _{_{0}}(\xi )+\Lambda _{1}\mathcal{G}(\xi ;\sqrt{2}%
)+V_{0}\mathcal{F}(\xi ,\alpha )  \notag \\
&&+\Lambda _{12}\mathcal{G}(\xi ;\sqrt{1+l_{r}^{2}})~,  \label{inte}
\end{eqnarray}%
where $\mathcal{F}(x;\gamma )$ is reported in Ref.~\onlinecite{Trallero3}
and $\mathcal{G}(x;z)$ is defined in the Appendix B. For the chemical
potential$,$ $\mu _{per,2},$ and the order parameter for the second species,
$\psi _{per,2},$ we obtain similar expressions by just changing $%
1\Leftrightarrow 2$ and $l_{r}\Leftrightarrow 1/l_{r}$ in Eqs.~(\ref{perturb}%
)$~$-$~$(\ref{inte}).

\section{Discussion of the results and conclusions.}

\begin{figure}[tbp]
\includegraphics[width = 0.48\textwidth]{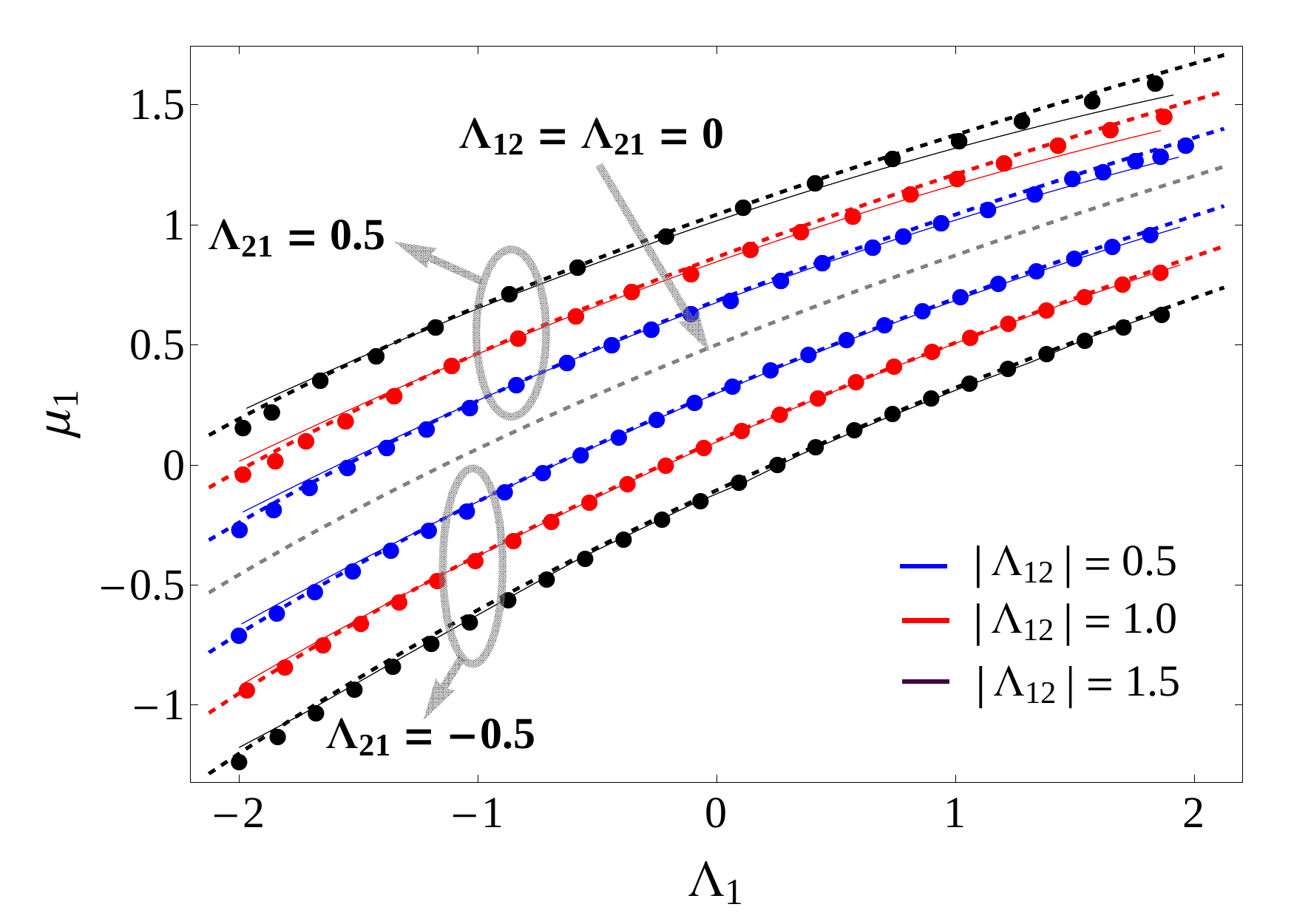}
\caption{(Color online) Dependence of the reduced chemical potential $%
\protect\mu_{per,1}=\protect\nu_{1}/\hbar\protect\omega$ on the
dimensionless self-interaction parameter $\Lambda_1$ for the inter-particle
term $\Lambda_{12}=\pm0.5, \pm1.0, \pm1.5$ and $\Lambda_{21}=\pm0.5$. Values
of $V_0=0$, $l_r=1$ and $\Lambda_{2}=1$ are fixed. Dashed and solid lines
represent the analytical results from Eqs.~(\protect\ref{muapr1}) and (%
\protect\ref{perturb}), respectively. Symbols correspond to the numerical
solution of Eq.~(\protect\ref{AdimCoup}). For sake of comparison, the limit
of one component ($\Lambda_{3}=0$) using Eq.~(\protect\ref{muapr1}) is
shown. }
\label{fig:fig1}
\end{figure}

In the following we present our results and discuss the reliability of the
two implemented methods of solution. It will be useful to compare the
obtained analytical expressions with direct numerical calculations. This
comparison allows to find ranges of values of the parameters $\Lambda _{1},$
$\Lambda _{2},$ $\Lambda _{12}$ $\Lambda _{21} $ and $V_{0}$ where the
variational approach and perturbation method can be implemented for the
description and predictions of the properties of the cigar-shape 1D
two-species Bose-Einstein condensates. For the numerical evaluation of the
system (\ref{AdimCoupIntegral}) we choose a finite difference method
described in Ref.~\onlinecite{Trallero3}.

\subsection{Chemical potentials}

\begin{figure}[tbp]
\includegraphics[width = 0.48\textwidth]{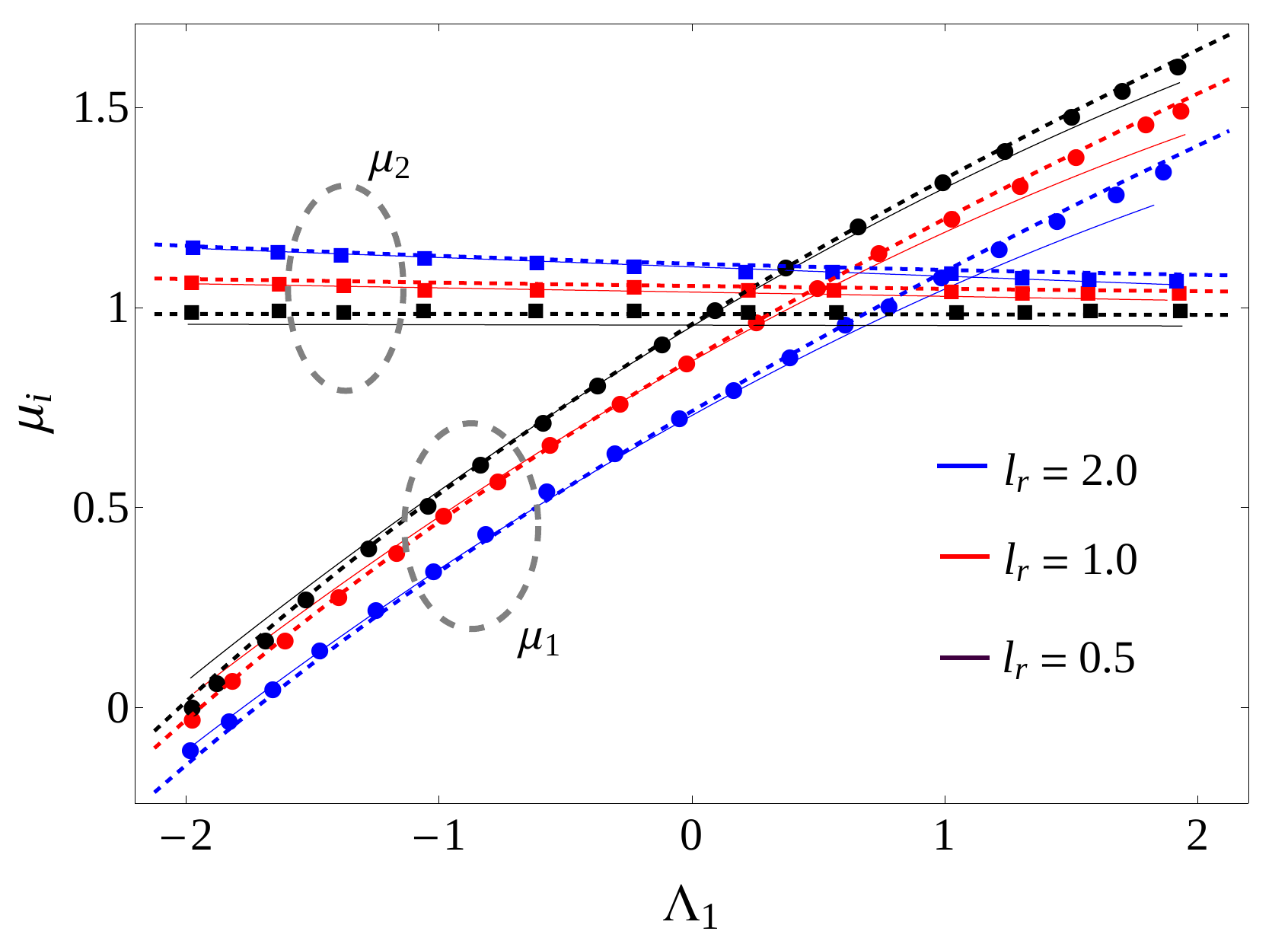}
\caption{(Color online) Dimensionless chemical potentials $\protect\mu _{1}$
and $\protect\mu _{2}$ as a function of $\Lambda _{1}$ for several species ($%
l_{r}=0.5$, 1.0 and 2.0). The same nomenclature as in Fig.~(\protect\ref%
{fig:fig1}) are employed. In the calculation we sorted $V_{0}=0$, $\Lambda
_{2}=1$, $\Lambda _{12}=1$ and $\Lambda _{21}=0.5$.}
\label{fig:fig2}
\end{figure}
First, we analyze the case when the intensity of optical lattice is turned
off, $V_{0}=0$. Figure~(\ref{fig:fig1}) shows the reduced chemical potential
$\mu_{1}$ as a function of the dimensionless non-linear term $\Lambda _{1}$
for the following values of the inter-species $\Lambda _{12}=\pm 0.5,$ $\pm
1,$ and $\pm 1.5.$ In the calculation we have fixed $\Lambda _{2}=1,$ $%
\Lambda _{21}=\pm 0.5,$ and $l_{r}=1.$ Variational approach calculations
given by Eqs.~(\ref{muapr1}) and (\ref{muapr2}) are indicated by dashed
lines, while the perturbation approach, using Eq.~(\ref{perturb}), is
symbolized by solid lines. Symbols represent the results obtained by direct
numerical evaluation of Eq.~(\ref{AdimCoup}). Taking as reference the
particular limit of one component, where $\Lambda _{3}=0$, as it is shown in
Fig.~\ref{fig:fig1}, we observe that the influence of the inter-specie
interaction on the chemical potential is to increase $\mu _{1}$ as the term $%
\Lambda _{12}>0$ increases, while the opposite result is achieved, i.e.,
$\mu _{1}$ decreases if $\Lambda _{12}<0$ decreases.

The small difference seen in the figure between the perturbation theory with
respect to the variational and numerical solutions for $\Lambda _{1}>0$ lies
in the range of validity of Eq.~(\ref{perturb}). In Ref.~%
\onlinecite{PhysicaD} it is shown that the perturbation theory for one
component reproduces quite well the chemical potential if $\left\vert
\Lambda _{1}\right\vert \lesssim 2 $. In the present case, the inter-species
interaction plays the role as an effective non linear term given by $\Lambda
_{1}$ $\left\vert \psi _{1}\right\vert ^{2}+\Lambda _{12}$ $\left\vert \psi
_{2}\right\vert ^{2}.$ Hence, the range of validity of Eq.~(\ref{perturb})
as function of $\Lambda _{1}>0$ is reduced if $\Lambda _{12}>0$. The
opposite we can argue if $\Lambda _{12}<0$, i.e., the function $%
\mu_{per,1}(\Lambda _{1})$ given by (\ref{perturb}) match the variational
and numerical calculations in a large range of values of $\Lambda _{1}>0$.
Similar arguments can be performed for the various combination of values of
the parameters considered in Fig.~\ref{fig:fig1}.

In Fig.~\ref{fig:fig2} we checked the influence of several species, $%
l_{r}=0.5$, 1, and 2, on $\mu _{1}$ and $\mu _{2}$ as function of $\Lambda
_{1}$ without optical lattice, $\Lambda _{2}=1$, $\Lambda _{21}=0.5$, and $%
\Lambda _{12}=1$. As might be expected, the chemical potential $\mu _{2}$ is
almost constant as a function of the self-interaction term of the first
species $\Lambda _{1}$. We note that for $l_{r}>1$ the value of the chemical
potential $\mu _{1}$ ($\mu _{2}$) is reduced (increased), while the opposite
it is obtained if $l_{r}<1$. This result is explained by the fact that the
effective inter-species $\lambda _{3}\left\vert \psi _{i}\right\vert ^{2}$
depends on the mass ratio $l_{r}$ (see Eqs.~(\ref{AdimCoup}), (\ref{muapr1}%
), (\ref{muapr2}) and (\ref{perturb})).

It can be seen that the variational approach fits very well the numerical
calculations, but the perturbation theory presents some differences as $%
\Lambda _{1}>0$ ($\Lambda _{1}<0$) increases (decreases). The same argument,
as it is given in the analysis of Fig.~\ref{fig:fig1}, we can argue for the
dependence of $\mu _{per,i}$ on $\Lambda _{1}$ and $l_r$. Nevertheless, this
analysis has to be taken with caution. The presence of the functions $f(z)$
and $g(z)$ in Eq.~(\ref{perturb}) establishes different ranges of validity
for $\mu _{i}(\Lambda _{1})$ as a function of $l_r$. Notice, that $f(z)<0$
for $z>0$, while $g(z)<0$ $(g(z)>0)$ for $z>1$ ($z<1$) (see Appendix B).
\begin{figure}[tbp]
\includegraphics[width = 0.48\textwidth]{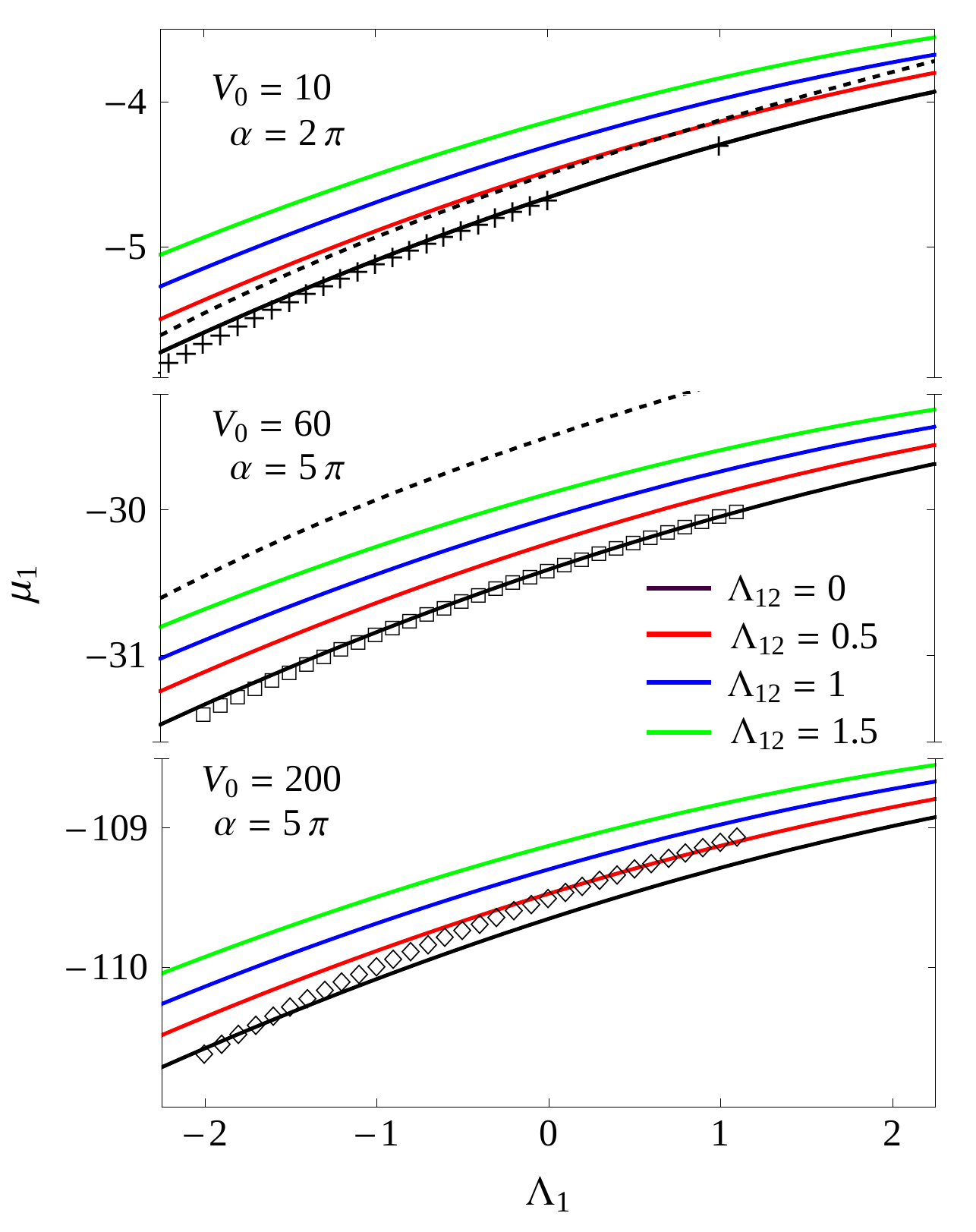} .
\caption{(Color online) The same as Fig.~\protect\ref{fig:fig1} for several
values of the reduced optical lattice intensity species ($V_{0}=10,\protect%
\alpha =2\protect\pi $ and $V_{0}=60,200$ with $\protect\alpha =5\protect\pi
$). The influence of the interspecies interaction is represented by solid
lines. Symbols are the numerical solution of the Eq.~(\protect\ref{AdimCoup}%
) and dashed lines the variation calculation using Eq.~(\protect\ref{muapr1}%
). $\Lambda _{2}=1$ and $\Lambda _{21}=0.5$.}
\label{fig:fig3a}
\end{figure}

\subsection{Influence of the optical lattice}

In Fig.~\ref{fig:fig3a} it is shown the behavior of the chemical potential
as function of $\Lambda _{1}$ for several values of the laser intensity $%
V_{0}$, the reduced wavelength $\alpha $ and the $\Lambda _{12}$ parameter.
Solid lines represent the calculation following Eq.~(\ref{perturb}), dashed
lines the variational approach as given by Eq.~(\ref{muapr1}) with $\Lambda
_{12}=0$. Symbols correspond to the numerical solution of Eq.~(\ref{AdimCoup}%
) for $\Lambda _{12}=0$. From Fig.~\ref{fig:fig3a} it can be seen that Eq.~(%
\ref{muapr1})\ does not match with the perturbation calculations neither
numerical solutions. As $V_{0}$ increases, the variational approach becomes
worse, reflecting the choice of the trial functions (\ref{TrialFunc}) we
have employed to calculate the energy. In connection with the perturbation
theory, the agreement is satisfactory for any $V_{0}$ less than 200, where a
small deviation from the exact numerical results is achieved. As it is
expected, the influence of the unlike interspecies interaction is to
increase the chemical potential (the opposite is obtained if $\Lambda
_{12}<0,$ not shown in the figure).

\subsection{Miscibility of the two species}

\begin{figure}[tbp]
\includegraphics[width = 0.48\textwidth]{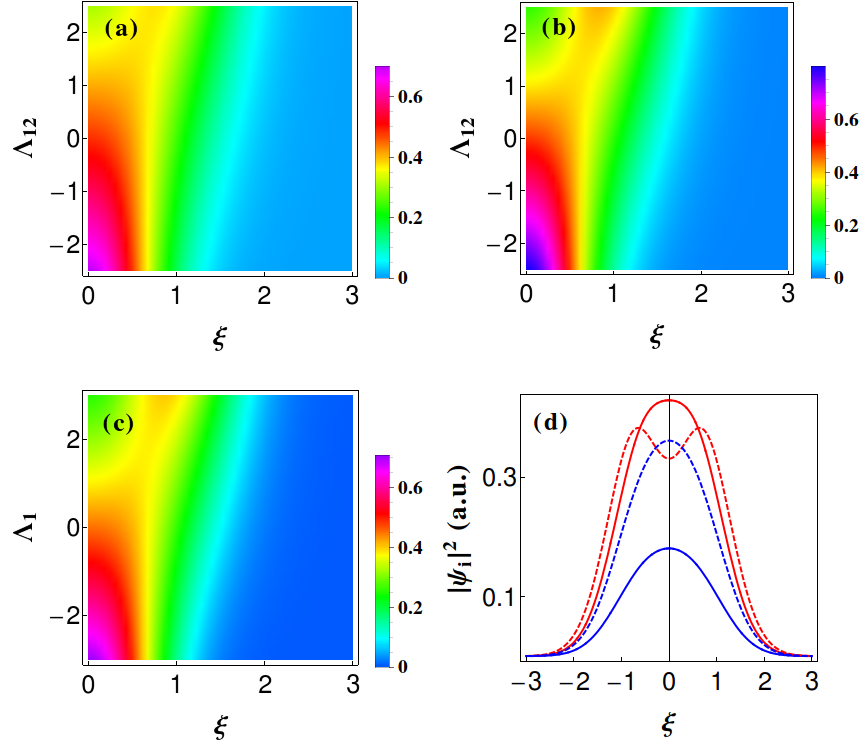} .
\caption{(Color online) Density profile of the species BEC for $V_{0}=0$.
Panel (a): As a function of $\Lambda _{12}$ for $\Lambda _{1}=1$ and $%
l_{r}=1 $. Panel (b): The same as panel (a) for $l_{r}=1.5$ Panel (c):
Varying $\Lambda _{1}$ for $\Lambda _{12}=1$, and $l_{r}=1.5$. Panel (d): {%
Functions $|\protect\psi _{i}(\protect\xi )|^{2}$, i=1(2) red (blue) for $%
\Lambda _{12}=0.5$ , $N_2/N1_=.8$ (solid line) and $\Lambda _{12}=0.8$, $%
N_2/N1_=.8 $ (dashed lines) 3. Here $\Lambda _{1}=1$ and $l_{r}=1.5$}}
\label{fig:fig4}
\end{figure}
A central issue for a description of the properties of multi species is the
evaluation of the order parameter as a function of particle-particle and
interspecies interaction. The control of the unlike particle interaction by
Feshbach resonance~\cite{4} allows to tune the miscibility or not of these
structures~\cite{8} and the challenge to create ultracold polar molecules.

Figure~\ref{fig:fig4} displays the spatial distribution density $\left\vert
\psi _{per,1}(\xi )\right\vert ^{2}$ as function of the dimensionless
parameters $\Lambda _{12}$ (panels (a) and (b)) and $\Lambda _{1}$ (panel
(c)). From Figs.~\ref{fig:fig4}(a) and (b) we observe the influence of one
species over another. The condensate is more delocalized as the
inter-species parameter $\Lambda _{12}$ increases. Also, as the mass of the
second species increases, the probability density $\left\vert \psi
_{per,1}(\xi )\right\vert ^{2}$ spreads on the space and the maximum of the
wavefunction decreases. The opposite is observed for the attractive
interaction when $\Lambda _{12}<0$, i.e., the density profile becomes more
confined at $\xi\approx0$ as $\Lambda _{12}$ decreases. Moreover, a stronger
localization occurs as the parameter $l_r$ increases. In other words, the
system with large mass difference presents a more effective attraction
between both components, which means that it
favors the miscibility among both species if $\Lambda _{12}<0$. A
comparison between attractive and repulsive dimensionless non-linear
parameter $\Lambda _{1}$ is sorted in panel (c) of the figure. As $\Lambda
_{1}$ increases from 0 to 3, the density is spread is space. Also, for $%
\Lambda _{1}$ large enough, the maximum of $\left\vert \psi _{per,1}(\xi
)\right\vert ^{2}$ is displaced by the particle-particle repulsive
interaction. In the case of attractive interaction, $\Lambda _{1}<0$, the
maximum of the order parameter $\psi _{per,1}(\xi )$ lies at the origin. {%
For sake of clarity,} in panel (d) we show the influence of the interaction $%
\Lambda _{12}$ {on the density profile $\left\vert \psi _{per,i}(\xi
)\right\vert ^{2}$ (i=1.2)}. Notice that the ground state is modulated by
the repulsive interaction induced by the species 2 and the maximum of
density probability is shifted to $\xi \neq 0 $ as $\Lambda _{12}$
increases. From the physical point of view this results are clear, the
species 2 is expelled off the origin by the first condensate. The mutual
repulsion between the two-species affect the spatial localization of density
profile {As we stated above}, this effect is driven not only by the values
of $\Lambda _{12}$, but also by the ratio of the masses involved in the two
condensates (see Eq.~(\ref{inte})).

The density distributions results of Fig.~\ref{fig:fig4} indicate in a
general way the degree of the immiscibility or phase separation of binary
condensate due to the interspecies repulsion. In our case the structure is
symmetric and it is related with the ratio of number of particles $%
N_{2}/N_{1}.$ These results are in complete concordance with recent
experimental reported observations for the $^{87}$Rb - $^{133}$Cs binary condensates.~%
\cite{5} The trial wavefunctions (\ref{TrialFunc}) cannot take into account
these behaviors over the spatial distribution as a function of $\Lambda
_{12} $, since they are \textsl{a priori\/} located at the origin.

In conclusion, we have derived simple explicit expressions for the chemical
potentials and order parameters in the case of two species
of non-homogeneous BEC, where the system is loaded in a harmonic trap
potential. We generalize the variational method for the case of two coupled
GP equations, showing that the obtained closed analytical expressions
for $\mu_{i}$ $(i=1,2)$ represent very good solutions for any values of the vector $%
\mathbf{\Lambda }$ if $V_{0}=0.$ Also, employing the
perturbation theory we are able to get analytical solutions for $\mu _{i}$
and the order parameter components $\psi _{i}$ as functions of the dimensionless
vector $\mathbf{\Lambda }$. By comparison with the numerical solutions we
found the range of validity of the Eq.~(\ref{perturb}). By the calculations
we show the strong dependence of $\mu _{i}$ and $\psi _{i}$ on the strengths $%
\Lambda _{1},$ $\Lambda _{2},$ $\Lambda _{12},$ $\Lambda _{21}$ and $V_{0}$.
This study gives a very useful result establishing the universal range where
each solution can be easily implemented. In particular, the dependence of the
order parameter $\psi _{i}$ on $\Lambda _{i}$ and $\Lambda _{ij}$ $(i\neq j)$
allows to study the immiscibility of two given species. We should note
that the variational model here developed can be extended to a cubic-quintic
model\cite{Trallero0} and allows to explore the
influence of quintic nonlinear terms on the ideal 1D two coupled pure cigar-like shape
system.

\section*{Acknowledgements}
This work was partially supported by the UFRJ and SECITI-DF/CLAF.
C~T-G.~wishes to acknowledge the hospitality of the Instituto de Matem\'atica, UFRJ.

\appendix

\section{Matrix elements}

The fourth dimensional matrix $\mathbf{S(}l_{r}\mathbf{)}$ is defined as

\begin{eqnarray}
S_{mn;pl}(l_{r}) &=&\frac{1}{\pi \sqrt{2^{n+m+l+p}n!m!l!p!}}\times  \notag \\
&&\int_{-\infty }^{\infty }\left[ \exp \left[ -(1+l_{r}^{2})z^{2}\right]
H_{n}(z)\right. ,  \notag \\
&&\left. H_{m}(z)H_{l}(l_{r}z)H_{p}(l_{r}z)\right] dz~,  \label{A1}
\end{eqnarray}%
with $H_{n}(z)$ the Hermitian polynomials~\cite{Abramowitz}$.$The matrix
elements $S_{mn;pl}(l_{r})$ have the followings properties:

i) $l_{r}S_{mn;pl}(l_{r})=S_{pl;mn}(1/l_{r});$

ii) $S_{2m0;00}(l_{r}),$ $S_{k0;0m}(l_{r})$ and $S_{km;00}(l_{r})$ are equal
to~\cite{Gradshteyn80}

\begin{equation}
S_{2m0;00}(l_{r})=\frac{(-1)^{m}\sqrt{(2m)!}}{\sqrt{\pi (1+l_{r}^{2})}2^{m}m!%
}\frac{l_{r}^{2m}}{(1+l_{r}^{2})^{m}}~,
\end{equation}

\begin{equation*}
S_{k0;0m}(l_{r})=\frac{\left( -1\right) ^{\frac{3m+k}{2}}\left( k+m\right) !%
}{\sqrt{\pi }\sqrt{k!m!}2^{\frac{m+k}{2}}\left( \frac{k+m}{2}\right) !}\frac{%
l_{r}^{k}}{\left( 1+l_{r}^{2}\right) ^{\frac{m+k+1}{2}}}~,
\end{equation*}

\begin{eqnarray}
S_{km;00}(l_{r}) &=&\frac{\left( -1\right) ^{\frac{k+m}{2}}2^{\frac{m+k}{2}}%
}{\pi \sqrt{k!m!}}\frac{\Gamma \left( \frac{k+m+1}{2}\right) l_{r}^{k+m}}{%
\left( 1+l_{r}^{2}\right) ^{\frac{m+k+1}{2}}}  \notag \\
&&F(-k-m,\frac{1-k-m}{2};\frac{1+l_{r}^{2}}{2l_{r}^{2}})~,
\end{eqnarray}%
with $\Gamma \left( z\right) $ the Gamma function~\cite{Abramowitz} and $F$($%
\alpha ,\beta ;z)$ the confluent hypergeometric function.~\cite{Gradshteyn80}

Using the above relations it is possible to get Eqs.~(\ref{perturb})-(\ref%
{inte}).

\section{Functions}

The functions introduced in Eq.~(\ref{perturb}) are defines as:
\begin{eqnarray}
f(z) &=&\ln \left[ \frac{1}{2}+\frac{1}{2}\sqrt{\frac{2+z^{2}}{2\left(
1+z^{2}\right) }}\right] ,  \label{F1} \\
g(z) &=&\ln \left[ \frac{1}{2}+\frac{1}{2}\sqrt{\frac{1+2z^{2}}{\left(
1+z^{2}\right) ^{2}}}\right]  \label{G1}
\end{eqnarray}%
\begin{equation}
h(z)=Ei(z)-\mathcal{C}-\ln z\text{\ \ ; \ }ch(z)=Chi(z)-\mathcal{C}-\ln z,
\label{h1}
\end{equation}%
where $Ei(z)=%
\mathchoice{{\setbox0=\hbox{$\displaystyle{\textstyle
-}{\int}$}\vcenter{\hbox{$\textstyle
-$}}\kern-.5\wd0}}{{\setbox0=\hbox{$\textstyle{\scriptstyle
-}{\int}$}\vcenter{\hbox{$\scriptstyle
-$}}\kern-.5\wd0}}{{\setbox0=\hbox{$\scriptstyle{\scriptscriptstyle
-}{\int}$}\vcenter{\hbox{$\scriptscriptstyle
-$}}\kern-.5\wd0}}{{\setbox0=\hbox{$\scriptscriptstyle{\scriptscriptstyle
-}{\int}$}\vcenter{\hbox{$\scriptscriptstyle -$}}\kern-.5\wd0}}%
\!\int_{-\infty }^{z}\frac{\exp (x)}{x}\,dx$ is the exponential integral, $%
Chi(z)$ the cosine hyperbolic integral, and $\mathcal{C}$ the Euler's
constant.

In Eq.~(\ref{inte}) the $\mathcal{G}(\xi ;z)$ is given by

\begin{equation}
\mathcal{G}(\xi ;z)=\frac{\exp (-\xi ^{2}/2)}{z\sqrt{\pi }\sqrt{\pi ^{1/2}}}%
\int\limits_{1}^{1/z}\frac{\exp \left[ -\frac{\xi ^{2}}{y^{2}}\left(
1-y^{2}\right) \right] -1}{1-y^{2}}dy.
\end{equation}

\end{document}